\documentclass[epj]{svjour}

\usepackage{graphics}

\begin{document}
\newcommand{\beq}{\begin{equation}}
\newcommand{\enq}{\end{equation}}
\newcommand{\lmax}{l_{\rm max}}
\newcommand{\lini}{l_{\rm i}}
\newcommand{\gat}{\Gamma_{\rm at}}
\newcommand{\dele}{\Delta E_{\rm sc}}

\title{Cold collisions in strong laser fields: 
partial wave analysis of magnesium collisions}

\author{J. Piilo\inst{1}, E. Lundh\inst{2}, \and  K.-A. Suominen\inst{1}}

\institute{Department of Physics, University of Turku, FI-20014 Turun yliopisto, Finland \and Department of Physics, Ume{\aa} University, SE-90187 Ume{\aa}, Sweden}

\date{\today}

\date{Received: date / Revised version: date}
\abstract{
We have developed Monte Carlo wave function simulation schemes to study
cold collisions between magnesium atoms in a strong red-detuned laser field.
In order to address the strong-field problem, we extend the Monte Carlo wave function framework to include the partial wave structure of the three-dimensional system. The average heating rate due to radiative collisions is calculated with two different simulation schemes which are described in detail.
We show that the results of the two methods 
agree and give estimates for the radiative collision heating rate for $^{24}$Mg atoms in a magneto-optical trap based on the $^1$S$_0$-$^1$P$_1$ atomic laser cooling transition.
}
\authorrunning{J. Piilo et al.}
\titlerunning{Cold collisions in strong laser fields: 
partial wave analysis of magnesium collisions}
\maketitle

\section{Introduction}

Trapping and cooling of neutral atoms in magneto-optical traps has become an important tool in atomic physics for spectroscopic and fundamental studies since it was first introduced in late 1980's~\cite{Metcalf03}. In such traps atoms are cooled down to the milliKelvin region and below it (depending on the particular element), and subsequent evaporative cooling in magnetic traps has made it possible to reach the quantum degeneracy limit, where bosonic atoms tend to form Bose-Einstein condensates~\cite{BEC}, and fermionic ones demonstrate the exclusion principle by filling the Fermi sea~\cite{DeMarco99,Truscott01}. Despite the success in going colder and denser, the behavior of the atomic cloud in basic magneto-optical traps still remains only a partially understood problem. One of the challenges is to understand if and how the atom-atom interactions in the form of inelastic cold atomic collisions affect the thermodynamics of the trapped and cooled cloud of atoms. Studies of such collisions have, however, also a more fundamental role, because cold collisions in the presence of light challenge the traditional approaches of collision and scattering theory~\cite{Suominen96,Weiner99}.

In general laser cooled and trapped neutral atoms provide an excellent system to study the basic aspects of collision theory. The slow relative motion of atoms guarantee the validity of the Born-Oppenheimer approximation and one can consider the collision as a one-cycle vibration of a quasimolecule formed by the colliding atoms (we prefer the term "quasimolecule" to "molecule" because any binding of atoms to each other is only temporary). One can separate the relative motion of two atoms into radial and angular parts, and the number of involved partial waves is limited to only a few lowest ones. Especially for the main isotopes of the alkaline earth atoms, such as Mg, Sr and Ca, the missing nuclear spin leads to a single, non-degenerate atomic ground state, and the bosonic nature of these atoms forces an even parity to the partial waves (or, in the quasimolecule picture, to the ground state rotational quantum numbers). Naturally, in the limit of very low temperatures only the $s$-wave is allowed, and under suitable conditions the atoms can form a Bose-Einstein condensate~\cite{Pethick01}.

The simple picture given above becomes distorted if we add the presence of light. Firstly, on a single atom level the absorption of a photon, although needed for laser cooling, will eventually limit the temperatures that one can achieve (the recoil limit)~\cite{Comment1}. Thus ultracold temperatures can be obtained either by evaporative cooling or by using very sophisticated absorption-free techniques such as velocity-selective coherent population trapping~\cite{Bardou02}, or Raman cooling~\cite{Kasevich92}. Secondly, the laser can also excite the quasimolecule formed by the colliding atoms~\cite{Gallagher89,Julienne89,Julienne91}. The result of this process depends very much on the frequency and intensity of the laser light. In the basic laser cooling situation, however, the light field is usually near-resonant and strong. Unfortunately, this is also the region where theoretical treatment of the problem is most difficult, and where the assigning of experimentally observable signals to specific processes is hard. 

The main bulk effects arising from collisions in the standard laser cooling and trapping process in a magneto-optical trap are assumed to be a) the loss of atoms from the trap, and b) the heating of the atomic cloud. The first process has been studied in detail both experimentally and theoretically, but mostly at weak fields (in a typical experiment the cooling lasers are shut off and replaced by a highly controllable, weak probe laser); for reviews, see Refs.~\cite{Suominen96,Weiner99}. The second process has been almost completely ignored due to the difficulty of modelling it~\cite{Suominen98}, and to the difficulty of separating the collisional heating from other heating effects such as reabsorption of scattered photons~\cite{Hillenbrand94}. 

In this paper we build a dynamical description of collisional heating by a strong laser field in a three-dimensional two-atom system. Our tool is wave packet dynamics~\cite{Garraway95}, i.e., time-dependent description of the collision process. Spontaneous emission of photons in the semiclassical laser field approximation makes the two-atom system an open quantum system~\cite{Carmichael02}, and we use the Monte Carlo quantum jump method as a tool~\cite{Dalibard92}. We take the dominant magnesium isotope $^{24}$Mg as our study case, because the lack of hyperfine structure reduces the number of molecular states and thus allows a quantitative study~\cite{Machholm01}, where the complexity arises only from the partial waves. Also, laser cooling and trapping of magnesium has been achieved experimentally~\cite{Sengstock94,Madsen02}. 

Previously near-resonant collision studies have been limited to two-state models, and at best the partial waves have been treated independently (ignoring second-order couplings)~\cite{Machholm01}, but at strong fields this is not possible as several waves are coupled to each other. We have recently taken a step
to the direction of including the partial wave structure in a more realistic way to the description
of cold collisions in strong fields~\cite{Piilo04a}. Of course, it is not possible to include all partial
waves in a realistic study, so the partial wave manifold must be truncated.

In building the framework for our wave packet calculations we have noticed that the artificial truncation of the infinite sequence of partial waves must be done with care, otherwise it may lead to significant errors. Another problem in cold collisions is the setting of the initial state for the atom-atom system, as the quasimolecule states are strongly coupled even asymptotically. After outlining the background for inelastic processes and the properties of the magnesium system in Sec.~\ref{framework}, we discuss these problems and our solutions to them. In Sec.~\ref{simulations} we discuss how to connect our simulations with a heating rate. By performing the wave packet calculations in our framework, we show that it is possible to assign a heating rate to the system. We also study the intensity-dependence of this rate. These results are reported in Sec.~\ref{sec:results}, where we also show that the long-time behavior of the atom-atom system, i.e., accumulation of relative kinetic energy in several subsequent collisions, can be understood and predicted reasonably well from the short-time one-collision data. Finally, in Sec.~\ref{sec:conclusions}, we discuss our results and their consequences. 

\begin{figure}[th]
\centering
\scalebox{0.5}
{\includegraphics{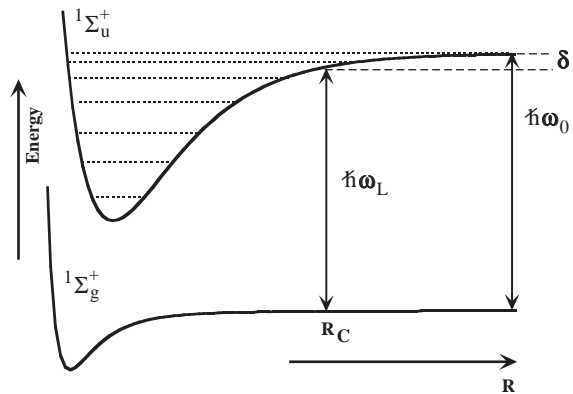}}
\caption[f1]{\label{schematics}
The quasimolecule view of the colliding magnesium atoms in a laser field.}
\end{figure}

\section{Framework}\label{framework}

\subsection{Inelastic processes}\label{frameworkA}

Let us consider the situation in Fig.~\ref{schematics}, where we show schematically the basic molecular potentials of our model, as a function of the atomic distance $R$. For best efficiency the frequency of a cooling laser ($\omega_L$) is usually set a few atomic electronic transition linewidths ($\Gamma_{\rm at}$ in energy units) below the exact resonance ($\omega_0$), i.e., detuned to the red side of the transition [detuning $\delta=\hbar(\omega_L-\omega_0)<0$]. A single atom is thus slightly off-resonant with the laser light. However, we can also view the situation in the quasimolecule picture. Then one of the molecular states can at some specific atomic distance $R_C$ (Condon point) couple {\it resonantly} to the molecular ground state. Usually there are many such states, but for $^{24}$Mg (and also for the other major isotopes of alkaline earth atoms) there are only two, namely ${\rm ^1\Sigma_u^+}$ and ${\rm ^1\Pi_g}$~\cite{Machholm01,Meath68,Stevens77,Czuchaj01}. Since the excitation of the 
${\rm ^1\Pi_g}$ state is strongly forbidden at long distances, the system reduces in practice to a simple system
consisting of two electronic states. Note that asymptotically the molecular ground state (${\rm ^1\Sigma_g^+}$ here) corresponds to both atoms being in the atomic ground state, and the excited molecular state corresponds to one atom being in the atomic ground state (here $^1$S$_0$) and the other one in the atomic excited state (here $^1$P$_1$). 

Eventually at very short atomic distances the potential for the excited molecular state becomes repulsive, and thus below the dissociation limit we have a set of bound vibrational states (dotted lines in Fig.~\ref{schematics}).
When the cooling laser is just below the excited state dissociation limit, the vibrational line broadening is usually too large for the states to be resolved. By increasing the laser detuning, the states become resolved and
one can perform spectroscopic studies of these states.
This leads to photoassociation spectroscopy~\cite{Weiner99}, which is a very useful tool for determining the $s$-wave scattering lengths for atomic collisions, and for creating translationally cold molecules from cold atoms. 

If we describe the situation in the dynamical picture based on only the electronic molecular states, the inability to resolve the vibrational states means simply that if the quasimolecule is excited at $R_C$, it has a high probability to decay back to the ground state via spontaneous emission of a photon, before reaching the small values of $R$. If the quasimolecule survives to the region of small $R$, its state may change into a one with lower energy (asymptotically), and the energy difference between these two excited states goes into an increase of the relative kinetic energy. This increase is large enough for the atoms to be lost from the trap, and the mechanism is called the fine-structure change (FS) for alkali metal atoms~\cite{Gallagher89,Julienne89,Julienne91}, and state change (SC) for alkaline earth atoms~\cite{Machholm01}. 

Another possibility is that the quasimolecule decays before reaching the small $R$ region, but has acquired enough relative kinetic energy for the atoms to exceed the trap depth (usually on the order of 1 K). In magneto-optical traps Doppler cooling works efficiently on hot atoms, but if the trap depth is exceeded, even this cooling can not recapture the hot atoms produced in collisions. Both effects lead to a non-exponential decay in the number of trapped atoms and can be observed via monitoring the number of atoms (collisions with background atoms etc. lead to exponential decay). This is called radiative escape (RE)~\cite{Holland94}.

Radiative heating, which is studied in this paper, follows the radiative escape mechanism, but occurs for atoms that have not managed to exceed the trap depth in their kinetic energy increase before decay. The loss mechanisms are very visible and important, but compared to them the radiative heating is really a bulk effect, especially at strong fields. The resonant excitation at $R_C$ is very strong, but the decay also concentrates to its vicinity. At strong fields the population that decays near $R_C$ can also undergo re-excitation~\cite{Suominen98}, i.e., we can not separate the dynamics into a sequence of separated excitation and decay processes, as has been done in the weak field studies. The inclusion of partial waves complicates the situation further, as we shall discuss in Sec.~\ref{frameworkE}.

\subsection{Magnesium quasimolecule}\label{frameworkB}

\begin{figure}[th]
   \centering
   \scalebox{0.5}
   {\includegraphics{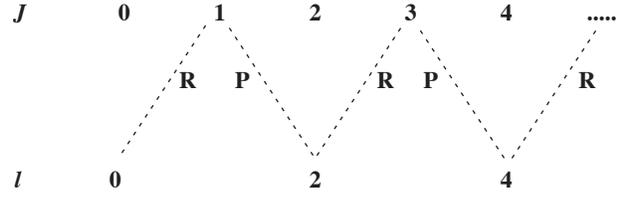}}
   \caption[f2]{\label{fig:Levels}
Partial wave $l$ and rotational state $J$ coupling scheme for the ground $^1\Sigma_g^+$ and excited $^1\Sigma_u^+$ states.}
\end{figure}

We have selected as our study case the $^1$S$_0$-$^1$P$_1$ laser cooling transition of $^{24}$Mg. The methods and qualitative conclusions apply to other alkaline earth atoms, and to a lesser degree also to other elements where, however, hyperfine
structure will complicate the picture. For the system of two $^{24}$Mg atoms, the ground electronic quasimolecule state is ${\rm ^1\Sigma_g^+}$ and, as discussed above, for excited state we choose the ${\rm ^1\Sigma_u^+}$ state, for which the P and R rotational state transition branches are possible, see Fig.~\ref{fig:Levels}. Since for symmetry reasons only even partial waves exist for the ground state, only the odd ones contribute in the excited state due to the missing Q branch.

For small detunings the Condon point $R_C$ lies in the long range where the ground
state is approximately flat, and the excited ${\rm ^1\Sigma_u^+}$ potential
is~\cite{Machholm01}
\begin{equation} \label{eq:V}
   U(R) = -\frac{3\gat}{2\left( k_r R\right) ^3}\left[\cos \left(k_r R\right) + 
          (k_r R) \sin \left(k_r R\right)\right],
\end{equation}
where $k_r$ is the wavenumber of the $^1$S$_0$-$^1$P$_1$ transition, $R$ is the internuclear distance, and $\gat$ is the corresponding atomic linewidth ($\gat/h=78.8$ MHz). For short range we actually flatten the excited state potential when it begins to correspond to kinetic increases that exceed the maximum values allowed by numerics (and survival to short range is small in any case). The position-dependent molecular linewidth including the relativistic retardation corrections is~\cite{Machholm01}
\begin{equation}\label{linewidth}
   \Gamma(R) = \gat\left\{ 1-\frac{3}{\left(k_r R\right)^3 } 
   \left[ (k_r R) \cos \left( k_r R\right) 
   -\sin \left( k_r R\right)\right]\right\}.
\end{equation}
The wave number $k_r$ forms the basis of the recoil unit system, which is used in the numerical computations. The recoil energy is $E_r=\hbar^2k_r^2/2\mu$, where $\mu=m_{\rm Mg}/2$ is the reduced mass of the two-atom system, and the recoil unit of length is $\lambda/2\pi=k_r^{-1}$, where $\lambda$ is the transition wavelength corresponding to $\hbar\omega_0$. For the $^1$S$_0$-$^1$P$_1$ transition of $^{24}$Mg, $E_R=k_B\times 9.8\ \mu$K and $\lambda=285.21$ nm~\cite{Machholm01}.

Within the rotating wave approximation we can shift the ground state up in energy by the photon energy $\hbar\omega_L$, giving the atomic states (and asymptotically for the molecular states) an energy difference equal to the detuning $\delta$. Thus the Hamiltonian of the system, with the matrix structure due to the partial waves made explicit, reads
\begin{equation}\label{hamiltonian}
   H = \left( 
   \begin{array}{ccccc}
	   \ddots & \vdots & \vdots & \vdots & \cdots \\
       \cdots & 
				U_{\rm kin} + U_{e,l-1}
				& V_{l-1,l} & 0 & \cdots\\
       \cdots & V_{l,l-1} & 
      U_{\rm kin} +U_{g,l} + \delta
       & V_{l,l+1} & \cdots\\
       \cdots & 0 & V_{l+1,l} &
      U_{\rm kin}+ U_{e,l+1}
       & \cdots\\
       \cdots&\vdots&\vdots&\vdots&\ddots
   \end{array}
   \right),
\end{equation}
where $U_{\rm kin}$ is the kinetic energy
\begin{equation}
U_{\rm kin} = -\frac{\hbar^2}{2\mu}\frac{\partial^2}{\partial R^2},
\end{equation}
and the potentials $U_g$ and $U_e$ include the atomic interaction potential $U(R)$ from
Eq.~(\ref{eq:V}) and the centrifugal part:
\begin{eqnarray}
   U_{g,l} &=& \frac{\hbar^2l(l+1)}{2\mu R^2}, \nonumber \\
   U_{e,j} &=& U(R) + \frac{\hbar^2j(j+1)}{2\mu R^2}.
   \label{eq:centrifugal}
\end{eqnarray}
The couplings between the ground state partial waves ($l$) and excited state rotational states ($j$) are defined as
\beq
   V_{jl} = \Omega \alpha_{jl}\sqrt{\frac{\Gamma}{\gat}},
\enq
where the factors $\alpha_{jl}$ are given in Table~\ref{tab:Couplings}, and $\Omega$ is the Rabi coupling in energy units. Following Ref.~\cite{Machholm01}, we can relate the Rabi coupling to laser intensity as
\beq
   \frac{\Omega}{\gat} = 0.5304\sqrt{I(W/cm^2)},
\enq
where $I$ is the laser intensity. The previous weak field work~\cite{Machholm01} was performed at the intensity
1 mW/cm$^2$, where\-as our work concentrates on values $\Omega/\gat\simeq 1$, indicating
the intensity region 1 W/cm$^2$, i.e., about three orders of magnitude stronger fields.

\begin{table}
\centering
\caption[t1]{\label{tab:Couplings}
Factors $\alpha_{jl}$ for the couplings between the various partial 
waves/rotational states. $l$ denotes the ground state partial wave angular momentum number and $j$ the corresponding excited rotational state. P denotes the branch where the partial wave $l$ couples to the excited rotational state with angular momentum $j=l-1$, and R denotes the branch coupling $l$ to $j=l+1$.}
\begin{tabular}{cccccccc}
\hline\noalign{\smallskip}
branch & $\alpha_{jl}$ ($l=0$) & $\alpha_{jl}$ ($l\neq0$) \\
\noalign{\smallskip}\hline\noalign{\smallskip}
P &  - & $\sqrt{\frac{l/3}{2l+1}}$ \\
R & $\sqrt{2/3}$ &  $\sqrt{\frac{\left(l+1\right)/3}{2l+1}}$ \\
\noalign{\smallskip}\hline
\end{tabular}
\end{table}


\subsection{Monte Carlo simulations}\label{frameworkC}

The Hamiltonian (\ref{hamiltonian}) is not sufficient for describing collisions in the presence of laser light, because it does not take into account the spontaneous decay of the excited state. At weak fields it is unlikely that the decayed population is re-excited, so one can add decay to the Hamiltonian~(\ref{hamiltonian}) as an imaginary potential $i\Gamma/2$ for the excited quasimolecule states, with the position-dependent $\Gamma$ given by Eq.~(\ref{linewidth}). The resulting problem can be solved with traditional time-independent scattering methods for trap loss~\cite{Suominen98,Machholm01}. For radiative heating we need to estimate also the kinetic energy change that took place before the decay, and in strong fields we also need to allow for re-excitation. This forces us to treat the problem as a time-dependent one, with initial and final states, rather than with boundary conditions for ingoing and outcoming single-energy quantum waves. In this picture, the colliding atoms are described by a wave packet of relative motion, initially peaked around some mean value of relative velocity $v$. When we refer to the initial velocity of the wave packet, it actually means $\langle v \rangle$; the same applies to the kinetic energy $E_k$.

Since the wave packet system is now an open quantum system, its description would require one to use a density matrix instead of a wave function. That would also double the spatial degrees of freedom [density matrix $\rho(R,R';t)$ vs. wave function $\Psi(R,t)$], taking the problem beyond current computational capacities~\cite{Garraway95}. However, we can nevertheless use the wave function description. In the Monte Carlo wave function scheme~\cite{Dalibard92}, the decay is incorporated in a stochastic manner. We add the $i\Gamma/2$ term to the Hamiltonian~(\ref{hamiltonian}). In addition, at each time step of duration $\delta t$ in the numerical propagation of the wave packet with time-dependent Schr{\"o}dinger equation, the wave packet is allowed to collapse onto the ground state with a probability 
\begin{equation}
   P_{\rm jump} = \delta t\sum_i\int dR [\Psi_e^i(R)]^*\Psi_e^i(R)\Gamma(R)/\hbar,
\end{equation}
where the sum goes over all excited quasimolecule states (here all the rotational states $j$). The occurrence of the collapse (a quantum jump) is determined by a random number. At the collapse $\Psi_g(R,t)$ is deleted, $\Psi_e(R,t)$ becomes the new $\Psi_g(R,t)$, and then the excited state component is set to zero~\cite{Comment2}. Whether a jump actually occurred or not, the normalization of the wave function is restored at each time step. 

The system properties are calculated as ensemble averages over several such random realizations. The benefit of the strong fields regime is that the excited state occupation is high, and jumps are frequent; this improves the statistical accuracy and the relevant properties, including the average kinetic energy as a function of time, can be obtained with reasonably small ensemble sizes. Typically our ensembles have 64 or 128 members, which is quite sufficient for calculating the bulk properties such as average kinetic energy~\cite{Suominen98,Holland94}. In cold collision studies this method has been applied previously to two-state models~\cite{Suominen98,Holland94},
to collisions in optical lattices~\cite{Piilo1,Piilo2}, and an initial step towards including the partial waves
to cold collision problems was presented in Ref.~\cite{Piilo04a}.

\subsection{Semiclassical picture}\label{sec:LZ}

The wave packet approach treats the molecular dynamics quantum mechanically. The laser field is assumed to be classical, though, so it appears as a dipole-mediated coupling of the electronic states. Decay via spontaneous emission is included with the quantum jump approach. Thus the reabsorption of a spontaneously emitted photon is ignored, but this does not really play a role in collision dynamics. Also, at the temperatures considered here, we can also ignore the photon recoil for absorption and emission events.

There is, however, an aspect of classical dynamics that one can add to the system. Initially the wave packet is rather well defined in position and momentum, so it is possible to define a classical collision trajectory for the approaching atoms. In other words, the excitation can be treated as a dynamical process where the quasimolecule traverses a level crossing at $R_C$ along a classical trajectory. If only two states are involved, and decay is ignored for the brief moment of passing $R_C$, one can apply the Landau-Zener model~\cite{Suominen96,Suominen98}. This model predicts an excitation probability
\beq\label{lzprobability}
   P_{\rm LZ} = 1 -\exp\left(-\frac{2\pi V_C^2}{\hbar v_C U'(R_C)}\right),
\enq
where $v_C$ is the relative momentum, $V_C$ is the coupling between the molecular states, and $U'(R_C)$ is the slope of the potential difference, all calculated at the crossing point $R_C$. 

Even with its limitations, this simple model enables one to define the concept of a "strong field". In the atomic case, a field is strong if one is at the saturation limit of the atomic transition. The on-resonance saturation intensity is usually defined as~\cite{Metcalf03} $I_s=\pi hc/(3\lambda^3\tau$), where $\tau$ is $\hbar/\gat$. For the magnesium $^1$S$_0$-$^1$P$_1$ transition one gets $I_s=0.444$ W/cm$^2$, which is of the same order as the intensities used in our simulations. But one should note that the quasimolecular excitation probability can be close to unity even when the atomic excitation is far from being saturated. The Landau-Zener expression indicates that slow motion or gentle potential slope at $R_C$ can compensate in situations where the ratio $V/\delta$ is otherwise small. Therefore the concept "strong field" must always be conditional to temperature, detuning and potential structure.

Without the partial waves, the magnesium situation would be an excellent application case for the two-state model. Now the question is, how well this approximation translates to strong fields and partial waves.

\subsection{Partial waves}\label{frameworkE}

\begin{figure}[th]
\centering
\scalebox{0.4}
{\includegraphics{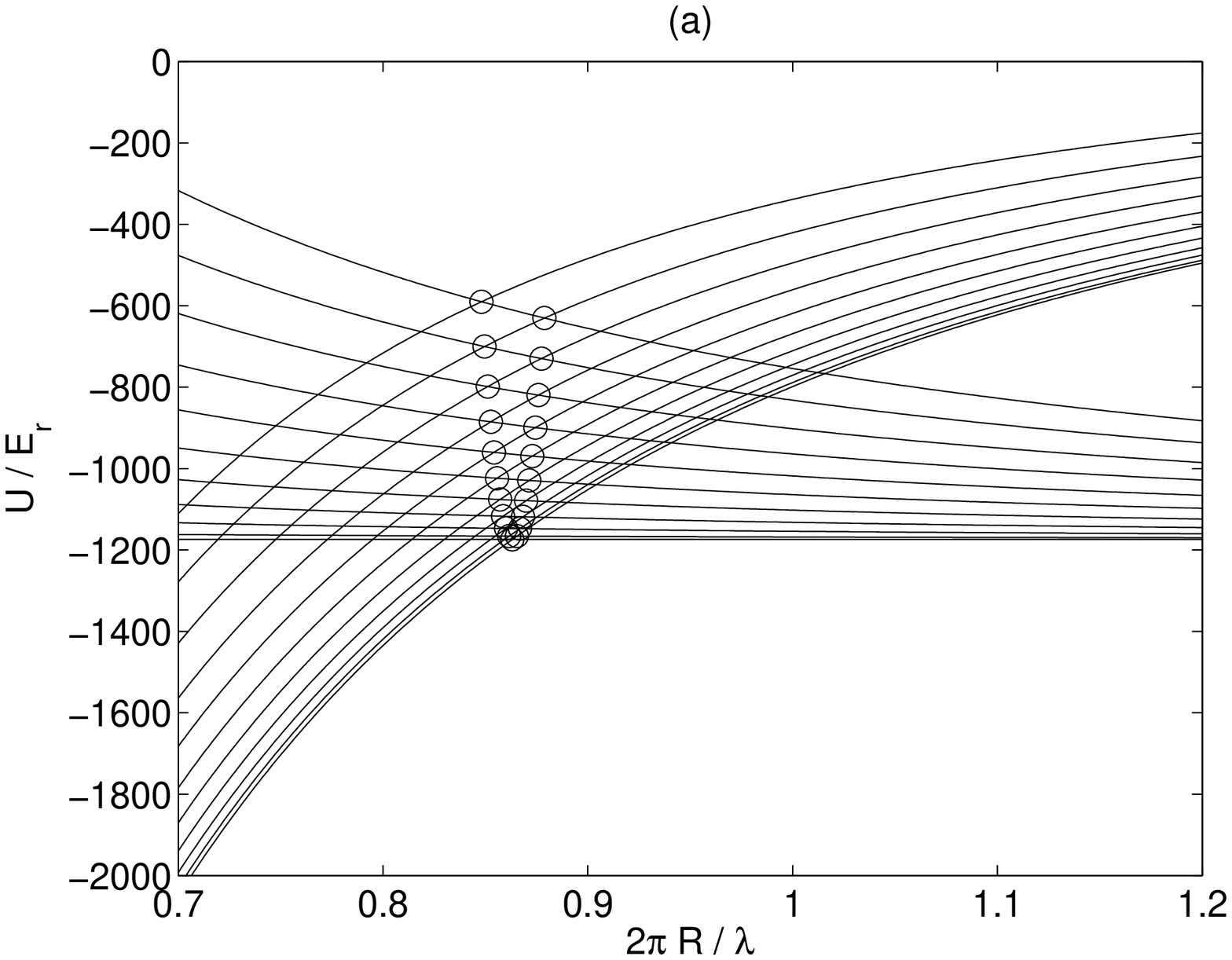}}
\scalebox{0.4}
{\includegraphics{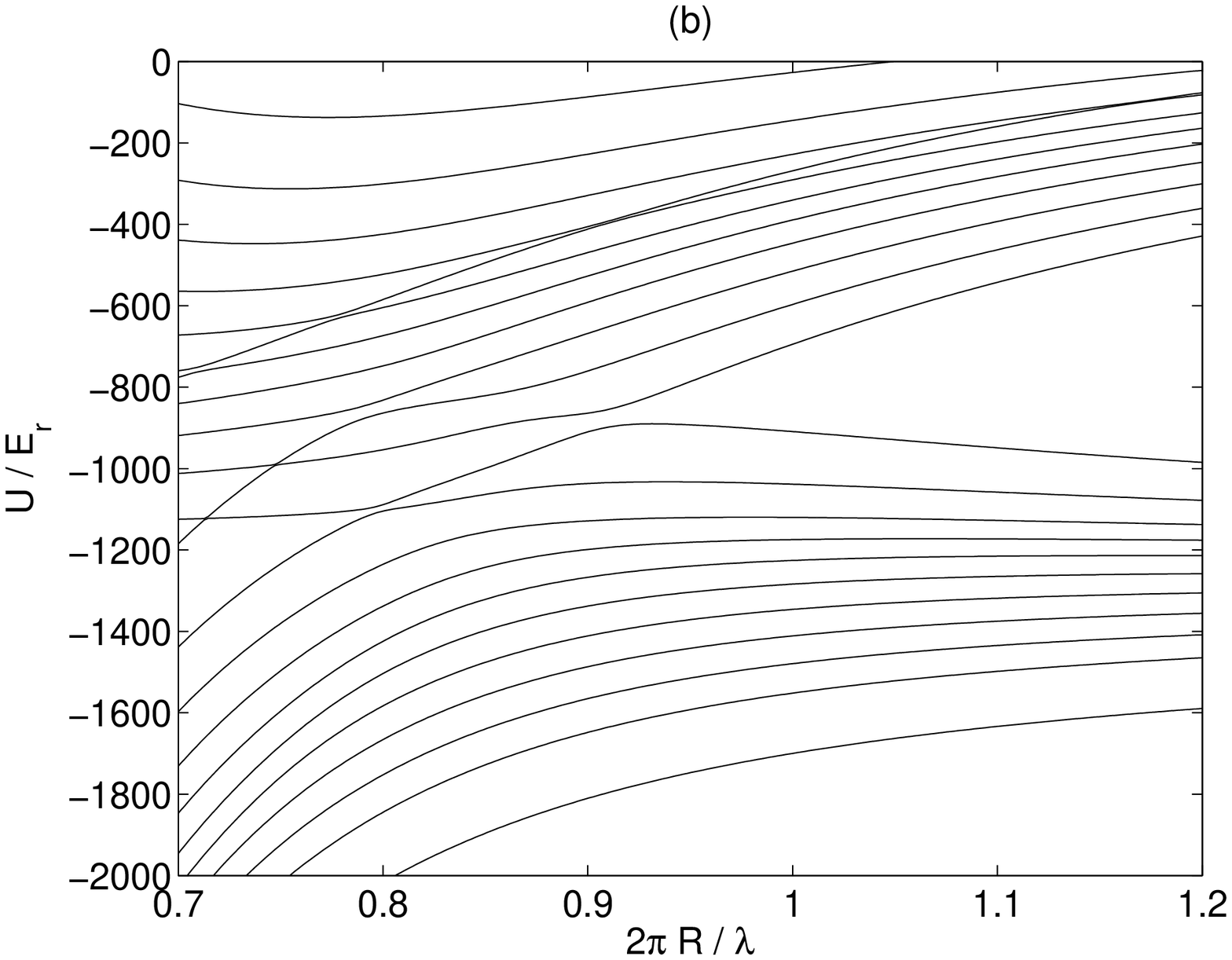}}
\caption[f3]{\label{fig:dressed}
(a) Ground and excited state potentials for partial waves/rotational states
up to $l=20$ and $j=21$. Open circles indicate the point of couplings forming a net of Condon points.
(b) The corresponding dressed potentials.The parameters are $\delta=-3\gat$ and $\Omega=1\gat$.}
\end{figure}

An example of the potential structure near the Condon point is shown in
Fig.~\ref{fig:dressed}(a). The couplings between the states form in fact a net of closely placed Condon points. The spatial separation between the individual two-state resonances suggests the possibility to partition the problem into separate two-state processes, which is exactly the approach adopted in weak field studies~\cite{Machholm01}. In this picture the approaching wave packet meets a sequence of two-state level crossings. It is then tempting to apply the Landau-Zener model to each crossing independently.

In our specific magnesium case the separation into a sequence of two-state models would have a qualitatively important aspect. Because the $Q$ branch is missing, there is a possibility for a "channelling effect". Now each ground state partial wave (with the exception of the $s$-wave) is coupled to two excited rotational states, but at different Condon points. Assume that the coupling $V_{jl}$ is strong enough that $P_{\rm LZ}$ is almost unity for all crossings. Any part of the wave packet that arrives on a ground-state partial wave with $l=l'>0$ will thus move to the excited state $j=l'-1$ at the first crossing. But in the next crossing, the excited rotational state is strongly coupled to the ground state partial wave with $l=j-1=l'-2$ (see Fig.~\ref{fig:Levels}). Thus the wave packet returns quickly to the ground state, where it climbs the centrifugal barrier, gets reflected, and follows the same path $l=l'-2\rightarrow j=l'-1\rightarrow l=l'$ backwards. This means that the collision is basically reduced to an elastic one for all partial waves except the $s$-wave.

At this point we note that in our simulations no such strong suppression was seen to occur. Basically, the two-state approximation fails because even at moderate fields the other states must be taken into account. This is best seen if one considers the dressed states, i.e., potential curves that correspond to the eigenstates of the coupled quasimolecule+light system, see Fig.~\ref{fig:dressed}(b). In order for the separation to independent two-state crossings to work out, the dressed states should show avoided crossings near each $R_C$. But it turns out that, for this to happen, the coupling strengths should be so low that $P_{\rm LZ}$  is far from unity. Therefore the division into separated, two-state Landau-Zener crossings is not applicable to the problem at hand, except for giving a rough tool to define the concept of strong field.

Strong fields correspond to the adiabatic limit, which means that the wave packet should follow the dressed states smoothly, and position-dependent unitary transformations describe the change from the bare state basis to the dressed state basis and back. The concept of adiabatic following and the basis change symmetry is broken by the spontaneous emission. It is still possible to treat the problem in the dressed state basis, but then the decay can go both ways as now practically {\it all} dressed states are superpositions of the ground and excited states. In the Monte Carlo method this leads to a complicated description, and therefore the bare states are usually preferred, even at strong fields. 

In the near-resonant situation the dressed and the field-free (bare) quasimolecule states do not match asymptotically at $R\rightarrow\infty$, which makes the setting of the initial conditions difficult. If we set the wave packet initially on one of the dressed states, which kinetic energy should we give it, the initial ground state one, or should we take into account the energy difference between the bare ground state, and the selected dressed state? Spontaneous emission  allows us to circumvent this problem. 

Since the atoms approach slowly, the quasimolecule reaches a steady state 
(in terms of the ground and excited state populations) before the wave packet
reaches the Condon point. The approximate time scale for steady state formation is a few times $\hbar/\gat$. It does not really matter whether we set the system on the bare or dressed state initially as it quickly adjusts itself to the steady state. Thus we can set the initial wave packet on a bare ground state corresponding to some particular value of $l$. It should be noted that in the Monte Carlo approach, the steady state appears only in the ensemble average. 

For practical reasons, we also need to truncate the infinite sets of partial waves and rotational states into finite ones, with some maximum values. Because of the centrifugal potential, only low-lying partial waves should contribute to a collision event. The maximum classically allowed angular momentum is obtained by matching the initial kinetic energy $E_k$ to the centrifugal potential at the Condon point $R_C$ (taken to be the {\it s}-wave Condon point, for simplicity), resulting in the value
\beq
   l_{\rm max,class} \simeq \frac{\sqrt{2\mu E_k}}{\hbar}R_C.
\enq
 
Excitation can occur also for higher partial waves than this classical value, because of the quantum nature of the dynamics. The wave packet has a finite width, i.e., there are
velocity components for which $E>E_k$. Also, there exists a kind of tunnelling effect for atoms that nearly reach $R_C$, which is quickly suppressed as $l$ increases. To be on the safe side, we have chosen $\lmax$ to be a few partial waves higher than the classical value; this has been found to be a satisfactory description in previous work~\cite{Holland94}, where the role of the partial wave structure was studied with wave packets assuming only independent Q branch ($l\rightarrow j=l$) processes.

It is important to avoid unwanted artificial effects that may arise because of the partial wave/rotational state truncation. In particular, choosing an odd number of partial waves may lead to the existence of a dark eigenstate in the Hamiltonian (\ref{hamiltonian}). Such a state is an almost pure superposition of ground-state levels and any population that occupies it will survive, unaffected by decay, and is not excited at the Condon point. If the system has a dark state, the frequent quantum jumps drive the quasimolecule population efficiently into it well before the actual collision occurs. 

Since a dark state can only exist if the Hamiltonian consists of an odd number of coupled levels, any suppression of radiative heating because of a dark state will be a pure artifact of the finite matrix and must be avoided. Indeed, our numerical computations indicate that in the case of few partial waves, the excited-state flux is much suppressed if the number of states is odd, but the difference between the odd and even cases is less pronounced when the number of levels is larger. All the results reported here are obtained using an even number of states. 

\section{Heating models}\label{simulations}

\subsection{Single-collision model}
\label{sec:singlecollision}

In order to obtain information on the heating rate we calculate the average kinetic energy increase due to radiative heating in a single collision (see Appendix A.1 for further details). In this model the initial wave packet is reasonably well defined in position and momentum, so that we can assign the initial relative velocity $v$ with the mean velocity of the packet, and the atoms are clearly separated. Next the atoms collide and heat up due to the radiative heating process, and move apart again. The single-collision simulation is stopped when the collision region is emptied of population and before the quasimolecule wave packet reaches the edge of the simulation space. From an ensemble of such simulations we obtain the average kinetic energy increase per collision $\dele$. 

\begin{figure}[t!]
\centering
\scalebox{0.4}
{\includegraphics{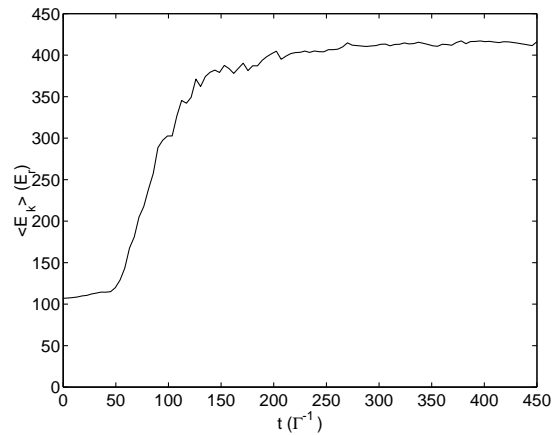}}
\caption[f3]{\label{fig:ExSingleCollision}
An example of the time evolution of the average kinetic energy in a single collision simulation (an ensemble of 64 realizations). The parameters are: $\delta=-3.0\gat$, $\Omega=1.0\gat$, $k_i=10.0k_r$, $\lmax=10$ and the initial component is $\lini=8$.}
\end{figure}

\begin{figure}[t!]
\centering
\scalebox{0.4}
{\includegraphics{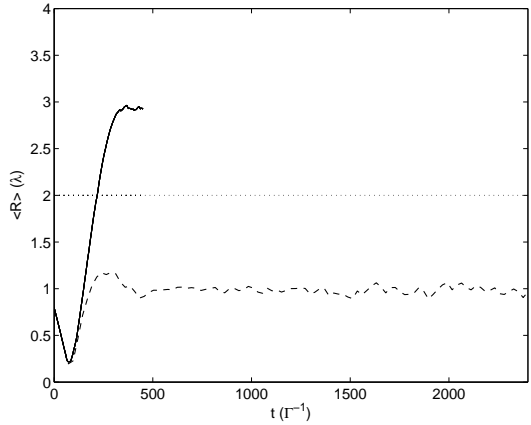}}
\caption[f3]{\label{fig:compare}
An example of the time evolution of the mean position of the wave packet, $\langle R\rangle$, in a simulation box of $L=4\lambda$ (solid line), and $L=2\lambda$ (dashed line). The parameters are: $\delta=-3.0\gat$, $\Omega=1.0\gat$, $k_i=10.0k_r$, $\lmax=10$ and the initial component is $\lini=8$.}
\end{figure}

This approach is conditional to the point that the post-collision behavior of the wave packet follows the above classical picture; an important aspect of our simulations is to show that this is indeed the case. Figure~\ref{fig:ExSingleCollision} shows an example of the time evolution of the average kinetic energy and we see a clear stepwise behavior of kinetic energy. Then $\dele$ can be calculated as the height of the energy step. The rough following of a classical trajectory is also demonstrated by the mean value of the wave packet position $R$ (solid line in Fig.~\ref{fig:compare}). Finally we note that the postcollisional distributions $|\Psi(R,t)|^2$ in all simulations clearly show that the collision region is emptied after the collision event. 

The inelastic process spreads the wave packet in momentum and thus in position, and eventually the fast parts get reflected first at the edge of the simulation space (at $R=4\lambda$ here), and $\langle R\rangle$ loses its classical-like behavior (Fig.~\ref{fig:compare}). Similarly, the wave packet is delocalized over the whole simulation space, and remains like that. This is reflected in the long-time behavior of $\langle R\rangle$ (dashed line in Fig.~\ref{fig:compare}). Clearly the classical-like trajectory description applies only to the first collision. For subsequent collisions we can not define any classical-like asymptotical initial and final states. 

Results from single-collision calculations are related to the physical heating rate coefficient of a thermal cloud at temperature $T$ through the relation 
(cf.~Appendix A.1)
\begin{eqnarray}
   K_H(T) =
   \frac{k_BT}{hQ_{T}} \int \frac{dE_k}{k_BT} e^{-E_k/k_BT}\times\nonumber\\
   \sum_{l (even)} (2l_i+1) \dele(E_k,l_i), \label{physrate}
\end{eqnarray}
where $Q_T = (2\pi\mu k_BT/h^2)^{3/2}$ is the translational partition function, and 
$\dele(E_k,l_i)$  is the numerical kinetic-energy increase calculated for the initial kinetic energy $E_k$ and initial ground state partial wave $l_i$. In practice we calculate the nonaveraged, energy-dependent rate coefficient
\begin{eqnarray}
   K_H(E_k) = \frac{E_k}{hQ_{E_k}}\sum_{l (even)} (2l_i+1) \dele(E_k,l_i).
   \label{nonaveragedrate}
\end{eqnarray}
The multi-level Monte Carlo computations are extremely time consuming, and we must therefore be content to calculate the nonaveraged rate coefficients $K_H(E_k)$ for only a few instances of parameter values. We argue that $K_H(E_k)$ still indicates the order of magnitude of the heating rate for temperatures around $T=E_k/k_B$.

For the weakly-coupled case that has been treated previously~\cite{Machholm01}, the wave packet populates the ground-state partial wave on which it was initially
concentrated, $\lini$, and in the collision region it populates also the 
neighboring excited-state levels, but is not allowed to spread beyond those. 
However, as the Rabi coupling $\Omega$ is increased, one quickly enters the regime where the population is distributed among several ground-state partial waves before the collision. The dependence on the initial state is therefore rather weak. For example, when $\Omega=1.0\gat$, the difference in kinetic-energy increase between the cases $\lini=0$ and $\lini=8$ is only 25\% (see Table~\ref{tab:SingleResults}). These dependencies are also affected by the choice of initial position, but the dependence of the final partial-wave-summed rates on the initial position appears to be weak.

\subsection{Multicollision model}\label{sec:Bathtub}

An alternative method to calculate the heating rate coefficient is a multicollision simulation. The essence of the method is that the quasimolecule population is reflected at the edge of the numerical space and thus performs repeated collisions. The beginning of the simulation resembles the single-collision situation (see Fig.~\ref{fig:compare}). But then multiple collisions begin to occur and the population is quite rapidly distributed over the whole simulation space. In a sense, the wave packet components slosh back and forth, spreading in momentum as they reach the Condon point, and spatial coherence is quickly lost. 

The quasimolecule gains kinetic energy when collisions occur, the wave packet moves faster, and the collision rate increases. We can now monitor the time evolution of the average kinetic energy and calculate the heating rate directly from the curve. The important point here is that the memory effects from initial conditions are lost and the result of the simulation should be independent of them. The quantum jump processes contribute strongly in mixing the population distribution of various partial waves between collisions. This brings the multicollision model closer to the actual situation where atoms emerging from a collision on some particular partial wave meet individually other atoms, forming a collision complex with partial waves that have nothing to do with the partial waves of the previous collision.

\begin{figure}[t!]
\centering
\scalebox{0.4}
{\includegraphics{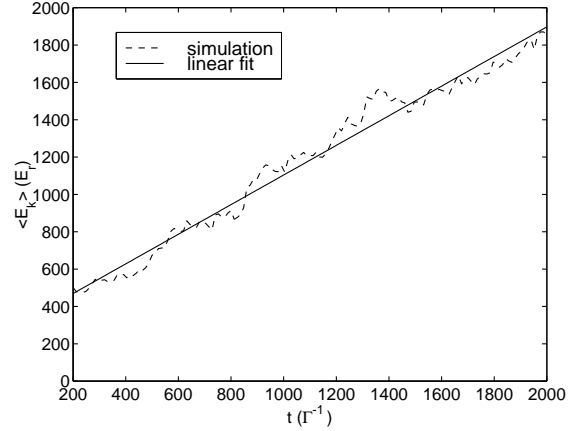}}
\caption[f3]{\label{fig:ExBathTub}
An example of the time evolution of average kinetic energy in the multicollision simulation. The kinetic energy behaves linearly and the slope of the curve gives the heating rate. Here $\delta=-3.0\gat$, $\Omega=1.0\gat$, $k_i=10.0k_r$, $\lmax=10$ and the initial component is $\lini=0$.}
\end{figure}

In the very beginning of a multicollision simulation we have the single-collision stepwise behavior in the average kinetic energy, but after the spatial coherence (i.e., localization to the vicinity of a classical trajectory) is lost, we find that the kinetic energy increases linearly, see Fig.~\ref{fig:ExBathTub}. This allows one to obtain the heating rate as the slope of the curve for the particular density corresponding to the size of the simulation space, as described in Appendix A.2. The increase in energy becomes non-linear later on. This is because the energy increase implies a faster collision rate and better survival, which compete with the weaker excitation probability (as discussed in Sec.~\ref{sec:LZ}). But we have found that at that point the atoms are already quite hot and strongly affected by Doppler cooling in any case. In order to obtain the heating rate in practice, we measure the slope of the energy in a time interval $[t_1,t_2]$, where the initial time $t_1$ is
chosen so that the wave packet has already had time to rid itself of
any dependence of initial state, and the upper limit $t_2$ is chosen well
before the curve becomes non-linear.

The numerical result from the multicollision computation is related to the actual
heating rate coefficient similarly to Eq.~(\ref{nonaveragedrate}), cf. Appendix A.2:
\begin{equation}
   K_H(E_k) =\frac{E_k}{hQ_{E_k}}\sum_{l (even)} (2l_i+1) \frac{dE_{\rm mul}}{dt}
   \frac{2L}{v}.
\end{equation}
Here, $L$ is the length of the simulation grid, $v$ is the
initial velocity $v=\sqrt{2E_k/\mu}$, and $dE_{\rm mul}$ is the energy of the wave packet
in the multicollision model.
Thus, the quantity under the
summation sign is the energy gain per collision (the factor 2 multiplying
$L$ reflects that the wave has to travel across the grid twice
between two collisions). Moreover, we have already noted that in the
multicollision model the results are more or less independent of the initial partial wave; thus, we directly perform the sum over even $l_i$ and obtain
\begin{equation}
   K_H(E_k) =
   \frac12(\lmax+1)^2\frac{\sqrt{2}\hbar^2 2L}{\mu E_k}
   \frac{dE_{\rm mul}}{dt}\simeq \sqrt{2}2LR_C^2\frac{dE_{\rm mul}}{dt}.
\label{finalr}
\end{equation}

As an internal check, the two simulation methods should be compared by matching the energy gain per collision:
\begin{equation}\label{modelcomparison}
   \dele =\frac{dE_{\rm mul}}{dt} \frac{2 L}{v}. 
\end{equation}
The left- and right-hand sides should be summed over partial waves before they can be compared, but anticipating the result that the single-collision simulations are almost independent of the initial angular momentum, the dependence can be neglected.

\section{Numerical results for heating rate coefficients}\label{sec:results}

The parameters for the simulations are chosen to correspond to realistic
experiments on magnesium. The atomic linewidth is $\gat = 391E_R$, and the recoil energy $E_R = k_B \times 9.8\ \mu$K. We have chosen to work with a detuning $\delta$ between 1 and 3 times the atomic linewidth $\gat$, and varied the Rabi coupling $\Omega$ between 0.1 and 2 times $\gat$, corresponding to laser intensities of 36 mW/cm$^2$ to 14.2 W/cm$^2$. The initial momentum $k_i$ has been chosen to be $k_i=10 k_r$ in most instances, giving an energy $E_k=100E_R=k_B\times 0.98$ mK. This is slightly below the Doppler temperature $T_D=1.9$ mK. We present the results of each of the two simulation methods, and compare them both to each other, and to the outcome of a simpler two-state model.

\subsection{Average kinetic energy increase}

\begin{figure}[t!]
\centering
\scalebox{0.4}
{\includegraphics {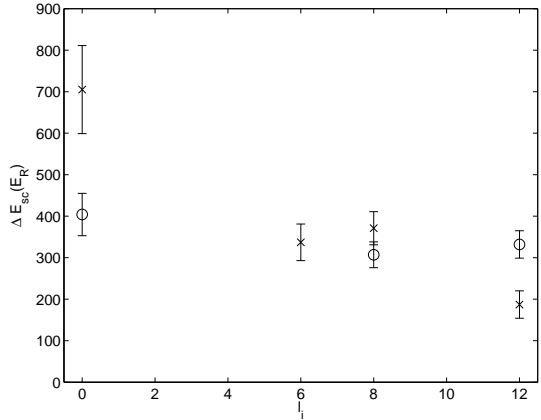}}
\caption[f3]{\label{fig:twotwelve}
Energy increase due to radiative heating in the single-collision model 
as a function of initial partial wave.
Crosses indicate the two-state result and circles are the results from using 
the full partial-wave structure (multistate model). The data refers to Table~\ref{tab:SingleResults}. 
}
\end{figure}

The results for the average kinetic energy increase in the single-collision model are displayed in Table~\ref{tab:SingleResults}. At the top of the table are the results from using only two partial waves, the ground-state partial wave as indicated in the column $\lini$ and the neighboring higher excited-state level, $j=\lini+1$. Below the horizontal bar are the results from the full calculation, incorporating the first 6 or 7 partial waves (i.e., up to $\lmax=10$ or 12). 
The initial mean position is set to $R_0=5\lambda/2\pi$. Some clear differences can be seen. First, when starting on the $s$-wave, the energy increase is larger in the two-state case than in the multistate case. The reason is rather trivial: in the two-state case there is only one channel for excitation; the probability for this
transition is proportional to 2/3, as can be seen in Table~\ref{tab:Couplings}. On the other hand, in the multistate calculation, many partial waves are occupied as the wave packet enters the crossing region, and the transitions to the excited state take place with a suppressed probability as can also be seen in
Table~\ref{tab:Couplings}; when $l$ increases towards infinity,
$\alpha_{jl}^2$ approaches $1/6$. Conversely, when $\lini$ is high
enough, the multistate result for the energy increase is higher 
because the population now spreads to lower partial waves before the collision. 
Figure~\ref{fig:twotwelve} illustrates this. These results demonstrate that in the
multistate case the redistribution of population before the collision
leads to a weak dependence on the initial partial wave, as already discussed in
Sec.~\ref{sec:singlecollision}.

The intensity dependence of the multistate single-colli\-sion
results is demonstrated 
in Table~\ref{tab:SingleResults2}, and we discuss it later, in Sec.~\ref{Sec:compare}.

\begin{table}
\centering
\caption[t1]{\label{tab:SingleResults}
The comparison of results from two-state and multistate single-collision models.
Here $N_l$ is the total number of states in the simulation (either 2 or 12), $\lini$ is the initial state, $k_i$ the initial momentum, and $\dele (E_R)$ the average increase of kinetic energy. The energy increase is calculated as a time average of the kinetic energy in the region where it is flat, subtracted by the initial energy. The error refers to statistical error of the kinetic energy at the
end of the Monte Carlo simulation.}
\begin{tabular}{ccccccc}
\hline\noalign{\smallskip}
$N_l$ & $\Omega (\gat)$ & $\delta (\gat)$ & $\lini$ & $k_{i} (k_r)$ & 
$\dele (E_R)$\\
\noalign{\smallskip}\hline\noalign{\smallskip}
2 & 1.0 & -3.0 &  0  & -10 & 705 $\pm$ 106 \\
2 & 1.0 & -3.0 &  6  & -10 & 337 $\pm$  44  \\
2 & 1.0 & -3.0 &  8  & -10 & 371 $\pm$ 40 \\
2 & 1.0 & -3.0 & 12 & -10 & 187 $\pm$  33  \\
\hline 
12 & 1.0 & -3.0 &  0  & -10 & 404 $\pm$ 51 \\
12 & 1.0 & -3.0 &  8  & -10 & 307 $\pm$ 31 \\
14 & 1.0 & -3.0 & 12  & -10 & 332 $\pm$ 33 \\
\noalign{\smallskip}\hline
\end{tabular}
\end{table}

\begin{table}
\centering
\caption[t1]{\label{tab:SingleResults2}
Intensity dependence of the results from the multistate single-collision simulations. The parameters are as in Table~\ref{tab:SingleResults}.}
\begin{tabular}{ccccccc}
\hline\noalign{\smallskip}
$N_l$ & $\Omega (\gat)$ & $\delta (\gat)$ & $\lini$ & $k_{i} (k_r)$ & 
$\dele (E_R)$\\
\noalign{\smallskip}\hline\noalign{\smallskip}


12 & 1.0 & -3.0 &  8  & -10 & 307 $\pm$ 31 \\
12 & 0.5 & -3.0 &  8  & -10 & 145 $\pm$ 16 \\
12 & 0.1 & -3.0 &  8  & -10 &   16 $\pm$   3 \\ 
\noalign{\smallskip}\hline
\end{tabular}
\end{table}

\subsection{Multicollision heating rate}

For the multicollision simulations, the size of the simulation box is set
to $L = 2.0\lambda$. Again, a two-state calculation
has been done in order to compare with the full partial-wave structure up to
$l_{\rm max} = 10$.
The results for the heating
rate are displayed in Table~\ref{tab:TubResults}.

Comparing the two- and twelve state results, there is a similar difference,
as already seen in the single-collision results. 
Moreover, and like discussed earlier, the single collision results indicated
only the weak dependance
of the initial state. We expect this to hold also especially
here for the multicollision model. Assuming that the initial state
dependence is weak to effect the first collision, the role
of the initial state should be even weaker for the following
collisions which actually determine the slope of the heating curve,i.e. the
heating
rate. This view is confirmed by the simulation result with $l_i=6$ in 
Table~\ref{tab:TubResults}.
Thus the results concentrate here on the intensity dependence of the heating
rate which is discussed in the following subsection.


\begin{table}
\centering
\caption[t1]{\label{tab:TubResults}
Comparison of the two-state and multistate results in the multicollision model. Here $\lini$ is the initial state, $k_i$ the initial momentum, and 
$dE_{\rm mul}/dt$ the slope of the energy-versus-time curve.}
\begin{tabular}{ccccccc}
\hline\noalign{\smallskip}
$N_l$ & $\Omega (\gat)$ & $\delta (\gat)$ & $\lini$ & $k_{i} (k_r)$
& $dE_{\rm mul}/dt (E_R\Gamma_{\rm at}/h)$  \\
\noalign{\smallskip}\hline\noalign{\smallskip}
2 & 1.0 & -3.0 &  0  & -10 &  4.48  \\
2 & 0.5 & -3.0 &  0  & -10 &  2.23  \\
2 & 0.4 & -3.0 &  0  & -10 &  1.00  \\
2 & 0.3 & -3.0 &  0  & -10 &  0.50  \\
2 & 0.2 & -3.0 &  0  & -10 &  0.22  \\
2 & 0.1 & -3.0 &  0  & -10 &  0.069 \\
\hline 
12 & 2.0 & -3.0 &  0  & -10 & 1.20   \\
12 & 1.0 & -3.0 &  0  & -10 & 0.79   \\
12 & 1.0 & -3.0 &  6  & -10 & 0.67   \\
12 & 0.5 & -3.0 &  0  & -10 & 0.40   \\
12 & 0.1 & -3.0 &  0  & -10 & 0.039 \\
\noalign{\smallskip}\hline
\end{tabular}
\end{table}

\subsection{Comparison of results}\label{Sec:compare}

The outcome of the single-collision and multicollision models are now to be
mapped to physical quantities using Eqs.~(\ref{nonaveragedrate}) and (\ref{finalr}). Equation~(\ref{nonaveragedrate}) is rewritten as 
\begin{eqnarray}
   K_H(E_k) =\left(\frac{\hbar^2\sqrt{2\pi}E_R^{1/2}}{\mu^{3/2}}\right)
   \frac12(\lmax+1)^2 \frac{\dele({\rm R.u.})}{k_{i}({\rm R.u.})}, \nonumber \\
   \label{factorizedrate}
\end{eqnarray}
where (R.u.) means recoil units. The fraction within brackets has the
numerical value 1.1307$\times 10^{-43}$Wm$^3$. The heating rates from the
multicollision computations are converted into energy increase per collision
with the conversion factor $2L/v=491$ in recoil units. The results are summarized in Table~\ref{tab:Physical}. The Monte Carlo computations are extremely time
consuming, but because of the weak dependence on initial partial wave and
the large error bars, it has not been necessary to perform the calculation for each $\lini$, but the values tabulated in Tables \ref{tab:SingleResults} and~\ref{tab:TubResults} have been sufficient. The correspondence between the
two models is seen to be fair.

\begin{figure}[t!]
\centering
\scalebox{0.4}
{\includegraphics{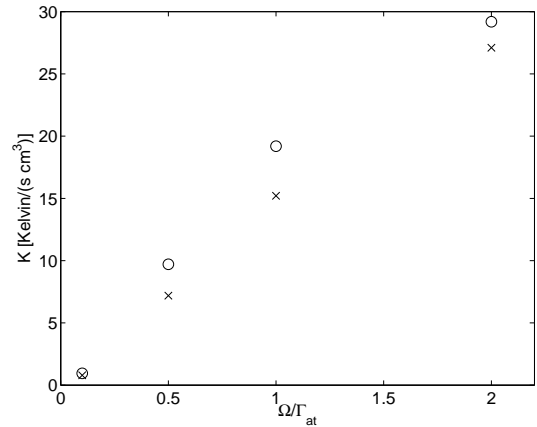}}
\caption[f3]{\label{fig:Rates}
Numerically calculated energy-dependent heating rate $R$ as a function of
Rabi coupling $\Omega$ for the energy $E_k=k_B\times 0.98$ mK and the
detuning $\delta=-3\gat$. Crosses are the results of the
single-collision simulations and circles are the results of the multicollision model.}
\end{figure}

If we assume $k_i=10k_r$, $\delta=-3.0\gat$, $V_C=2\Omega/3$, take the leading $R^{-3}$-term of the potential $U(R)$ and ignore the partial waves ($l=j=0$), the argument in the exponential in the Landau-Zener term in Eq.~(\ref{lzprobability}) becomes equal to $7.2\times (\Omega/\gat)^2$. For $V_C=\Omega/6$, the asymptotic limit for the $\alpha_{jl}$ for large $l$, this decreases to $1.8\times(\Omega/\gat)^2$. This means roughly
that in simulations with $\Omega=\gat$ we are more or less saturating the Landau-Zener two-state result for all partial waves. Although the model is not directly applicable to the multistate case, this result indicates nevertheless, that we should be in a region where the usual weak-field expectation of linear increase of the excitation rate with intensity $I$ fails.  Figure~\ref{fig:Rates} plots the heating rate as a function of the Rabi coupling for $\delta=-3\gat$ and $k_i=-10k_r$. We see an increase which is at first roughly linear in $\Omega$, but then becomes even weaker, indicating a possible saturation.  This is in qualitative agreement with two-state models~\cite{Suominen98}; when the Landau-Zener process saturates, the re-excitation of decayed population leads to an increase in excited state survival, which is linear in $\Omega$~\cite{Comment3}. 

\begin{table}
\centering
\caption[t5]{\label{tab:Physical}
Physical heating rates in units of $10^{-40}$Wm$^3$ given by the two
numerical methods.
$R_{s}$ stands for single collision and $R_{mul}$ for multicollision
heating rates. All the results are computed in the twelve-state model
allowing for the full partial-wave structure.}
\begin{tabular}{cccccccc}
\hline\noalign{\smallskip}
$\Omega (\gat)$ & $\delta (\gat)$ & $E_k/k_B$ (mK) & 
$R_s$ & $R_{mul}$\\
\noalign{\smallskip}\hline\noalign{\smallskip}
2.0 & -3.0 & 0.98 & 3.74 & 4.03 \\
1.0 & -3.0 & 0.98 & 2.10  & 2.65 \\
0.5 & -3.0 & 0.98 & 0.99 & 1.34 \\
0.1 & -3.0 & 0.98 & 0.11 & 0.13 \\
1.0 & -3.0 & 3.92 & 1.21 & - \\
0.5 & -1.0 & 0.98 & - & 0.23 \\
\noalign{\smallskip}\hline
\end{tabular}
\end{table}

\section{Discussion and conclusions}\label{sec:conclusions}

We have explored in detail the regime of strong laser fields in cold alkaline-earth collisions, which necessitates taking the full partial-wave structure into account.  For Rabi couplings $\Omega$ of the order of or larger than the linewidth $\gat$, the wave function of the colliding atoms quickly spreads and populates
all partial waves before and during the collision, yielding results that depend only weakly on initial conditions. 

The outcome of the two computational models, the ordinary simulation of a single collision and the multicollision model where the wave sloshes back and forth in a finite space, are seen to match well.
 {\it A priori} one tends to associate the multicollision model
as a more rigorous way to calculate the heating rate. This is because
consecutive collisions, during the system's long term
time evolution, keep track on the changes of the system
properties in a more
realistic way compared to the single collision model. 
However,  since the results produced by the two methods have reasonable
agreement, we come to the conclusion that the computationally
lighter single collision method provides a more practical approach.

It is worth noting that recent experiments on alkaline-earth atoms in 
magneto-optical traps show that  the Dop\-pler limit is not
reached~\cite{Xu02,Thomsen03,Kaiser05}. Our earlier results
indicate that the radiative heating does not prevent
reaching of the Doppler limit for current rather low atomic
densities in MOTs~\cite{Piilo04a}. The improved method
for cold collisions, multicollision model presented here,
gives further confirmation for this view. The authors of Ref.~\cite{Kaiser05}
attribute the extra heat to the transverse spatial intensity fluctuations
of their one dimensional optical molasses. However, a full 
three dimensional set-up may introduce additional factors to the problem.
From the point of view of atomic collisions, the partial wave treatment
includes three dimensional aspects, and the conclusions of this paper
remain the same.

In conclusion, we have developed methods,
based on Monte Carlo wave function simulations,
to include partial waves to the cold collision problems in strong laser fields. 
The methods have been described in detail highlighting
the most relevant aspects of the problem. For example, the appropriate
number of partial waves has to be even to avoid the artificial dark
state to influence the dynamics. The results
show the linear increase of the heating rate with laser intensity,
even in the regime where the Landau-Zener
excitation probability saturates. This is in qualitative 
agreement with two-state models when the re-excitation
of decayed population has been accounted for.
We think that the work presented here presents
a considerable step towards a complete partial wave treatment
of cold collisions in the presence of near-resonant
light. This topic is also relevant in the general framework
of fundamental collision theory.
A completely satisfactory and realistic model remains as a challenging task for
future studies.
 
\begin{acknowledgement}
The authors thank I.~Mazets and J.~Vigu\'{e} for useful comments, and M.~Machholm and P.~Julienne for discussions. The Finnish IT Centre for Science (CSC) is gratefully acknowledged for providing the supercomputer environment used in the simulations.
The work was supported by the Academy of Finland (projects 207614, 108699, 105740,
206108), the Magnus Ehrnrooth Foundation, the EU network CAUAC (Contract No.~HPRN-CT-2000-00165)
and the EU Transfer of Knowledge project CAMEL (Grant 
No. MTKD-CT-2004-014427).
\end{acknowledgement}

\section*{Appendix A: Rate coefficients}

\subsection*{A.1 Single collision rate}\label{AppA}

The collisional heating rate $\kappa_H (n)$ describes the kinetic energy change per unit volume and unit time. For indentical particles it will be equal to $K_Hn^2/2$, where $K_H$ is the rate coefficient for heating, and $n$ is the atomic density. The factor of $1/2$ removes the doubling in collision counting for identical particles. The total collisional heating rate is obtained by integrating $\kappa_H$ over trap volume. For simplicity we ignore the trap loss, so $n$ is constant in time. In any case, since the rate coefficient $K_H$ is density-independent, any time dependence in $n$ will not affect it.

For trap loss rate coefficient $K_{\rm loss}$ one normally calculates the collision frequency times the probability for loss, i.e., $v\sigma_{\rm loss}$, where $v$ is the relative velocity, and $\sigma_{\rm loss}$ is the cross-section for collisions, including the probability for a loss event to occur. This simple classical picture is connected to thermodynamics by assuming a distribution of velocities, $f(v)dv$, over which we take an average, and obtain $K_{\rm loss}=\langle v\sigma_{\rm loss}\rangle$. The dynamics of the two-body collision on the microscopic level then enters in calculating $\sigma_{\rm loss}$.

To calculate the heating rate $K_H$ is slightly more complicated, because in addition to the probability of an inelastic collision event to happen, we also need to estimate the amount of kinetic energy increase associated with it. Technically, we would have a continuous distribution of final energies corresponding to each value of $v$, due to the randomness of the spontaneous emission events. For practical reasons we consider an averaged rate, i.e., to calculate the average change in kinetic energy per unit time. Since we can perform the averaging over final energy states before the averaging over initial states, we define a heating cross section $\sigma_H(v)$ (units energy$\times$distance$^2$), that gives the difference of the average final relative kinetic energy and initial relative kinetic energy.

In the partial wave approximation we can write the quantum mechanical cross-section for identical atoms in a three-dimensional trap as~\cite{Julienne89}
\begin{equation}
   \sigma(v) =\frac{\pi}{k^2}\sum_0^{\infty} (2l+1)P_l(v),
\end{equation}
where $k$ is the wave number related to $v$ and $P_l(v)$ is the event probability. Thus, as a generalization, we write for heating
\begin{equation}
   \sigma_H(v) =\frac{\pi}{k^2}\sum_0^{\infty} (2l+1)\Delta E_{\rm sc}(v,l),
\end{equation}
where $\Delta E_{\rm sc}(v,l)$ is the average single-collision energy increase related to the "initial" partial wave $l$. 

In an isotropic Maxwell-Boltzmann gas the distribution of relative velocities $v$ is given by
\begin{equation}
   f(v)dv=4\pi\left(\frac{\mu}{2\pi k_B T}\right)^{3/2}v^2 
   e^{-\mu v^2/(2 k_B T)}dv,
\end{equation}
where $k_B$ is the Boltzmann coefficient, $\mu$ is the reduced two-particle mass (equal to $m_{\rm at}/2$ for identical atoms), and $T$ is the gas temperature at equilibrium. By defining a dimensionless energy ratio $\varepsilon=\frac{1}{2}\mu v^2/(k_BT)$ we get
\begin{equation} 
   K_H(T) = \frac{k_B T}{hQ_T}\int_0^{\infty}d\varepsilon e^{-\varepsilon} 
   \sum_0^{\infty} (2l+1) \Delta E_{\rm sc}(\varepsilon,l).
\end{equation}
If we truncate the partial wave sum e.g. with the classical trajectory argument, then the upper limit for the partial wave sum becomes energy dependent. Therefore it is preferable to keep the sum inside the energy integral.

Due to the complexity of the Monte Carlo simulations we are, in practice, limited to calculating $\Delta E_{\rm sc}(v,l)$ for a rather narrow range of initial relative velocities $v$. As in Ref.~\cite{Machholm01}, we define a single-energy rate as
 \begin{equation} 
   K_H(\varepsilon) = \frac{k_B T_{\varepsilon}}{hQ_{T_{\varepsilon}}}
   \sum_0^{\infty} (2l+1) \Delta E_{\rm sc}(\varepsilon,l).
\end{equation}
As one can see immediately, this mimics the case where $\varepsilon=1$, and the energy dependence on the integrand comes only from the Boltzmann factor. 

\subsection*{A.2 Multicollision rate}\label{AppB}

The classical picture in the derivation of a rate coefficient is a particle with velocity $v$, travelling a distance $v\Delta t$ in time interval of length $\Delta t$, and having the "area" given by the cross section $\sigma$. With gas density $n$, it meets $v\Delta t\sigma n$ particles in that time, so that the collision frequency for one particle becomes $v\sigma n$. The density of such particles brings in another $n$.  

In the multicollision model the fixed simulation box size $L$ gives an alternative way to approach the problem. Now the effective collisional volume is $2L\sigma$, in which the multiple collisions eventually become inseparable, and the kinetic energy does not increase in a stepwise manner. Here the quantum jumps and steady state formation make sure that apart from the change in the kinetic energy, there is no memory effects between collisions. Therefore, on average the situation corresponds to multiple inelastic collisions when we have a particle density $1/(2L\sigma)$, although in reality the particles do not collide at regular intervals. If we normalise the rate of change to density, we get
\begin{equation}
   K_H(\varepsilon) = \frac{k_B T_{\varepsilon}}{hQ_{T_{\varepsilon}}}
   \sum_0^{\infty} (2l+1) \frac{dE}{dt} \frac{2L}{v}.
\end{equation}

There is one {\it caveat}, though. If the change in kinetic energy is a strong function of $v$, then clearly for suitably long times $\frac{dE}{dt}$ is not a constant, and one would need to revert to calculating heating rates from single-collision results. The fact that in simulations $\frac{dE}{dt}$ remains a constant for a wide energy region makes it possible to use results from multicollision studies in the manner descibed above. More importantly, it implies that as one approaches the Doppler cooling temperature, the heating rate per partial wave remains constant.


\begin{thebibliography}{99}
\bibitem{Comment1} The scattering rate of photons sets the Doppler limit, which is 2 mK for the strong cooling transition $^1$S$_0$-$^1$P$_1$ in magnesium, in comparison with the photon recoil limit, 10 $\mu$K. In alkali atoms the degeneracy of the ground state allows one to break the Doppler limit and reach the recoil limit with tools like Sisyphus cooling, but this will not work for the alkaline earth atoms that have no such degeneracy.
\bibitem{Comment2} In fact in simulations we also determine from which particular state $j$ the collapse originated, and along which branch it took place, so after a jump only one ground state partial wave is occupied.
\bibitem{Comment3} When the Landau-Zener process saturates and re-excitation dominates, we can assume the delayed decay model~\cite{Suominen98}, where the exponential decay starts only once the quasimolecule has reached the edge of the re-excitation zone. The location of this edge is roughly where $|U(R)-U(R_C)|\simeq \Omega$, so the energy increase obtained by the quasimolecule before the exponential decay is proportional to $\Omega$.
\end{thebibliography}
\end{document}